\documentclass{article}
\usepackage{CJK}   
\usepackage{indentfirst}  
\usepackage{bm}    
\usepackage{amsmath,cite}
\usepackage{graphicx}  

\begin{document}    

\begin{CJK*}{GBK}{song}  

\thispagestyle{empty} \vspace*{0.8cm}\hbox
to\textwidth{\vbox{\hfill Chin. Phys. B \textbf{22} (2013) 028701\hfill}}
\par\noindent\rule[3mm]{\textwidth}{0.2pt}\hspace*{-\textwidth}\noindent
\rule[2.5mm]{\textwidth}{0.2pt}


\begin{center}
\LARGE\bf Challenges in theoretical investigations on configurations of lipid membranes$^{*}$   
\end{center}

\footnotetext{\hspace*{-.45cm}\footnotesize $^*$Project supported by National
Natural Science Foundation of China (Grant No 11274046).}
\footnotetext{\hspace*{-.45cm}\footnotesize $^\dag$Corresponding author. E-mail:  tuzc@bnu.edu.cn}

\begin{center}
\rm Z. C. Tu$^{\rm a)\dagger\rm b)}$
\end{center}

\begin{center}
\begin{footnotesize} \sl
${}^{\rm a)}$ Department of Physics, Beijing Normal University, Beijing 100875, China\\   
${}^{\rm b)}$ Kavli Institute for Theoretical Physics China, CAS, Beijing 100190, China\\   
\end{footnotesize}
\end{center}

\vspace*{2mm}

\textbf{Abstract: }This review reports some key results in theoretical investigations on configurations
of lipid membranes and presents several challenges in this field which involve (i) exact
solutions to the shape equation of lipid vesicles; (ii) exact solutions to the governing
equations of open lipid membranes; (iii) neck condition of two-phase vesicles in the budding state;
(iv) nonlocal theory of membrane elasticity; (v) relationship between symmetry and the magnitude of free energy.



\section{\label{sec-introd} Introduction}

Biological membranes are the basic elements of cells and cellular organelles. A membrane consists of a lipid bilayer mosaicked a variety of proteins. As model systems, lipid bilayer membranes are the leading research objects in the field of membrane biophysics \cite{KITPC2012}. Due to the large aspect ratio between the lateral dimension and thickness as well as the small compressibility, a lipid membrane is usually regarded as an incompressible elastic thin film in mechanics and a smooth surface in mathematics when we concern its large scale behaviors.
The geometry of the surface can be determined by its mean curvature and Gaussian curvature while the equilibrium configurations of membranes correspond to the local minimum of the free energy.
The bending energy contributes the most crucial effect on the free energy, which is usually taken as the Helfrich's form\cite{helfrich73}:
\begin{equation}\label{eq-helfrich}
f_H=\frac{k_c}{2}(2H + c_0)^2 + \bar{k}K,
\end{equation}
where $k_c$ and $\bar{k}$ are two bending rigidities. The former should
be positive, while the latter can be negative or positive for lipid membranes. $H$ and $K$ represent the local mean curvature and Gaussian curvature of the membrane surface, respectively. $c_0$ is called the spontaneous curvature which reflects the asymmetry of lipid distribution or other chemical or
physical factors between two leaves of lipid bilayers. Since the spontaneous curvature model can also be obtained from
symmetric argument for 2-dimensional (2D) isotropic
elastic entities, it is of generic significance
not only for lipid membranes, but also for other membranes consisting of 2D isotropic
materials such as carbon nanotubes and graphene\cite{Tu2008jctn}.

Based on Helfrich's spontaneous curvature model, the equilibrium configurations of lipid vesicles were deeply investigated in the past forty years\cite{LipowskyN91,Seifert97ap,OYbook1999,TuJGSP2011,Mladenovctp13}.
In stead of fully presenting the previous theoretical advances in this field, we will propose five challenges according to these theoretical advances and the author's personal flavors in this review. Of course, when interpreting these challenges, we still briefly mention some theoretical advances. The rest of this review is organized as follows. In section 2, we present the shape equation to describe equilibrium configurations of lipid vesicles. Then we show some analytic solutions and their corresponding configurations including sphere, torus, biconcave discoid, and so on. It is a big challenge to find other solutions to the shape equation. In section 3, we present the governing equations to describe equilibrium configurations of the open lipid membranes and verify a theorem of non-existence. Here two challenges are respectively related to the minimal surfaces with boundary curve and neck condition of two-phase vesicles in the budding state. In section 4. we discuss the nonlocal theory of membrane elasticity which is beyond the Helfrich's model. We can still derive the governing equation to describe the configurations of vesicles. It is a big challenge to seek possible analytic solutions to the governing equation. In section 5, we investigate the relationship between symmetry and the magnitude of free energy and argue that on what conditions the higher symmetric configurations correspond to lower free energy within the framework of Helfrich model. In the last section, we summarize the challenges again and call on physicists and mathematicians to overcome these challenges.

\section{Solutions to the shape equation of lipid vesicles}
Here we will discuss configurations of lipid vesicles composed of uniformly distributed lipids.

\subsection{Shape equation}
Since experiments have revealed that the area of lipid membrane
is almost incompressible and the membrane is impermeable for the
solutions in both sides of the membrane, the equilibrium configuration of lipid vesicle is expected to correspond to the local minimal of the extended Helfrich's free energy:
\begin{equation}\label{eq-exthelfrich}
F_H=\int\left[\frac{k_c}{2}(2H + c_0)^2 + \bar{k}K\right]\mathrm{d}A + \lambda A+ pV,
\end{equation}
where the integral is taken on the whole membrane surface of the vesicle. $A$ and $V$ represent the total area of the membrane surface and the volume enclosed in the vesicle, respectively. $\lambda $ and $p$ are two Lagrange multipliers which constrain the constant $A$ and $V$ when the vesicle takes various possible configurations. They can be physically understood as the apparent surface tension and osmotic pressure (pressure difference between the outside and inside) of the lipid vesicle.

Minimizing the free energy in the configuration space corresponds to the variational problem. The first order variation of the free energy (\ref{eq-exthelfrich}) leads to the shape equation\cite{OYPRL87,OYPRA87} of vesicle, which reads
\begin{equation}\tilde{p}-2\tilde{\lambda}
H+(2H+c_0)(2H^2-c_0H-2K)+\nabla^2(2H)=0\label{shape-closed}\end{equation}
with reduced parameters $\tilde{p}=p/k_c$ and $\tilde{\lambda}=\lambda/k_c$. In physics, this formula represents the force balance along the normal direction of membrane surface.

\begin{figure}[pth!]
\centerline{\includegraphics[width=8cm]{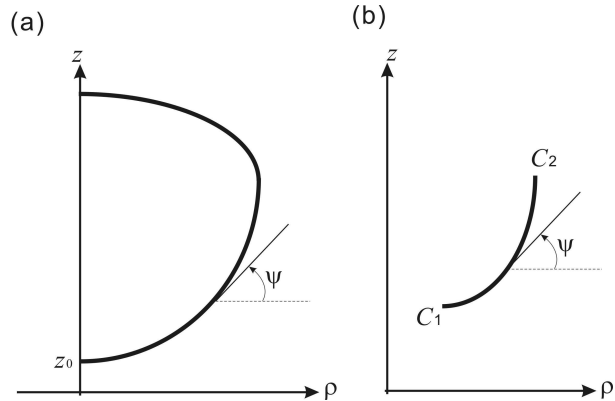}}
\caption{The generation curves for an axisymmetric vesicle (a) and an open membrane (b).\label{figgencuve}}
\end{figure}

Consider an axisymmetric vesicle generated by a planar curve shown in figure~\ref{figgencuve}a. $\psi$ is the angle between the tangent of the generation curve and the horizontal plane, with which the shape equation (\ref{shape-closed}) can be transformed into~\cite{HuOY93PRE,HuOY95CPB}
\begin{equation}
\tilde{p}+\tilde{\lambda}
h+(c_{0}-h)\left(\frac{h^{2}}{2}+\frac{c_{0}h}{2}-2K\right)-\frac{\cos \psi }{\rho}(\rho\cos \psi h')'=0,\label{shape-symmetr}\end{equation}
where $h\equiv {\sin \psi }/{\rho}+(\sin\psi)'$ and $K={\sin \psi
}(\sin\psi)'/{\rho}$. The `prime' represents the derivative with respect to $\rho$.
The shape equation (\ref{shape-symmetr}) of axisymmetric vesicles is a third-order differential
equation. Zheng and Liu \cite{zhengliu93} found the first integral $\eta_{0}$ for this equation and then
transformed it into a second-order differential equation
\begin{equation}\cos\psi h'
+(h-c_{0}) \sin\psi\psi^{\prime}-\tilde{\lambda} \tan\psi+\frac{2\eta_{0}-\tilde{p}\rho^2}{2\rho\cos\psi}-\frac{\tan\psi}
{2}(h-c_{0})^{2} =0.\label{firstintg}\end{equation} It is found that the present shape equation of axisymmetric vesicles degenerates into the form derived by Seifert \emph{et al.} \cite{SBLPRA91} when $\eta_{0}=0$ in equation~(\ref{firstintg}) which holds for vesicles with spherical topology free from singular points \cite{Podgornikpre95}.

\subsection{Typical solutions}

Up to date, we have known several analytic solutions to shape equations (\ref{shape-closed}) or (\ref{firstintg}). They correspond to surfaces of constant mean curvature (including sphere, cylinder, and unduloid), torus, biconcave discoid, unduloid-like surface and cylinder-like surfaces, and so on \cite{OYbook1999,NaitoPRL95,Konop97,MladenovEPJB02,GuvenPRE20022D,MladenovJPA08,Zhouxh2010,ZhangOYPRE96,Zhangcpb97}. Among them, only sphere, torus, and biconcave discoid are closed configurations which can be sketched as follows.

Firstly, let us consider a spherical surface with radius $R$. Then $H=-1/R$ and $K=1/R^2$. Substituting them into equation~(\ref{shape-closed}), we derive
\begin{equation}\tilde{p}R^2+2\tilde{\lambda}
R-c_0(2-c_0R)=0.\label{sphericalbilayer}\end{equation}
Under proper conditions, the parameters $c_0$, $\tilde{p}$, and $\tilde{\lambda}$ take proper values such that the solution to the above equation exists.

\begin{figure}[pth!]
\centerline{\includegraphics[width=8cm]{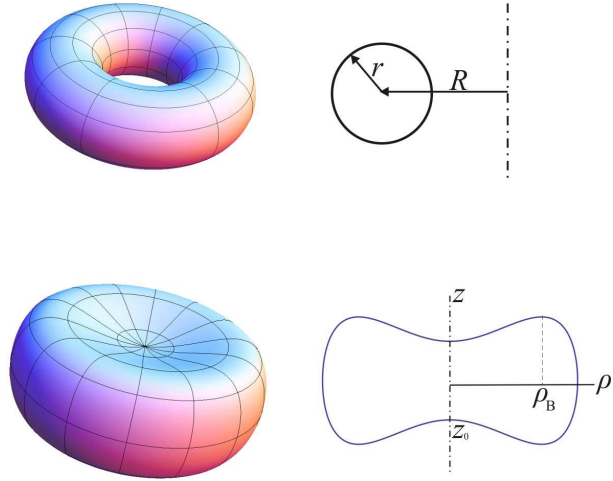}}
\caption{(color online) Torus and Biconcave discoid as well as their generation curves.\label{figbiconcv}}
\end{figure}

Secondly, a torus shown in figure~\ref{figbiconcv} is a revolution surface generated by a circle with radius
$r$ rotating around an axis in the same plane of the circle. The
revolving radius $R$ should be larger than $r$.
The generation curve can be expressed as\cite{OYbook1999,oypra90}
\begin{equation}
\sin\psi = (\rho/r) - (R/r).\label{toruseqsym}
\end{equation}
Substituting it into equation~(\ref{firstintg}), we arrive at $R/r=\sqrt{2}$, $2\tilde{\lambda} r=c_{0}( 4-c_{0}r)$,
$\tilde{p}r^{2}=-2 c_{0}$ and
$\eta_{0} = -1/r \neq 0$.

Thirdly, for $0<c_0\rho_B<\mathrm{e}$, the parameter
equation
\begin{equation}\left\{\begin{array}{l}\sin\psi=c_0\rho\ln(\rho/\rho_B)\\
z=z_0+\int_0^\rho \tan\psi d\rho
\end{array}\right.\label{solutionbicon}\end{equation}
corresponds to a planar curve shown in figure~\ref{figbiconcv}. Substituting it into equation~(\ref{firstintg}),
we have $\tilde{p}=0$, $\tilde{\lambda}=0$, and
$\eta_0=-2c_0 \neq 0$. That is, a biconcave
discoid generated by revolving this planar curve around
$z$-axis can satisfy the shape equation of vesicles. This result can give a good explanation to
the shape of human red blood cells under normal physiological
conditions \cite{NaitoPRE93,NaitoPRE96}.

It is necessary to point out that the inverted catenoid \cite{Castro07} is also a closed surface satisfying the shape equation. However, the poles of this surface contact tightly with each other, which is not permitted by real physical systems.

\subsection{Challenge}
Can we further find the other analytic solutions to the shape equation (\ref{shape-closed}) or (\ref{firstintg}) which represent the closed vesicles without self-contact? Under certain conditions, equation (\ref{firstintg}) can be extremely simplified. Considering $h= {\sin \psi }/{\rho}+(\sin\psi)'$, if we chose a new variable
\begin{equation}\xi = \frac{\sin \psi }{\rho}+\frac{\mathrm{d}(\sin\psi)}{\mathrm{d}\rho}-c_0,\label{eq-varxi}\end{equation} equation (\ref{firstintg}) can be expressed as a very concise form:
\begin{equation}\frac{\mathrm{d}}{\mathrm{d}\rho}\left(\frac{\cos\psi}{\xi}\right) + \frac{\tan\psi}{2}=0\label{eq-simpshe}\end{equation}
when $\tilde{\lambda}$, $\tilde{p}$ and $\eta_0$ are vanishing.
It might be much easier to find solutions to the above equations (\ref{eq-varxi}) and (\ref{eq-simpshe}) than the original shape equation. However, it is still a challenge to find the solutions to these equations.

On the other hand, we need to consider other probabilities if all our efforts are in vain.
Among all closed non-intersect surfaces, there are probably only sphere, torus and biconcave discoid that can satisfy the shape equation and can be expressed as the elementary functions. It is also valuable to make this negative proposition verifiable or falsifiable.

\section{Solutions to the governing equations of open lipid membranes}
Here we will discuss configurations of open lipid membranes composed of uniformly distributed lipids.

\begin{figure}[pth!]
\centerline{\includegraphics[width=7cm]{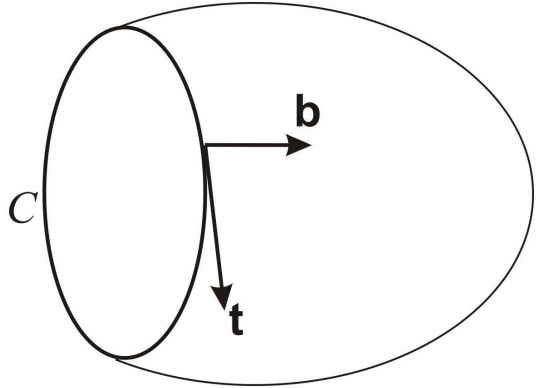}}
\caption{Open smooth surface with a boundary curve $C$. For the point in $C$, $\mathbf{t}$ and $\mathbf{b}$ are located in the tangent plane of the surface. The former is the tangent vector of $C$ while the latter is perpendicular to $\mathbf{t}$ and points to the side that the surface is located in. \label{figopenm}}
\end{figure}

\subsection{Governing equations}
As shown in figure~\ref{figopenm}, a lipid membrane with a free edge can be expressed as an open smooth surface with a boundary curve $C$ in geometry. Because the freely exposed edge is energetically unfavorable, we assign the line tension (energy cost per unit length) to be $\gamma>0$. Then the free energy that we need to minimize can be expressed as
\begin{equation}F_{O}= \int \left[ \frac{k_c}{2} (2H +c_0)^2 + \bar{k} K \right] \mathrm{d} A +\lambda A +\gamma L ,\label{eq-frenergyn2}\end{equation}
where $L$ is the total length of the free edge.

By using the variational method, the shape equation
\begin{equation}(2H+c_{0})(2H^{2}-c_{0}H-2K)-2\tilde\lambda H+\nabla
^{2}(2H) =0, \label{eq-openm}\end{equation}
and three boundary conditions
\begin{eqnarray}
&&\left[ (2H+c_{0})+\tilde{k}\kappa_n\right]_{C} =0,\label{bound1} \\
&&\left[ -2{\partial H}/{\partial\mathbf{b}}+\tilde\gamma
\kappa_n+\tilde{k}\dot{\tau}_g]\right] _{C} =0,\label{bound2}\\
&&\left[ (1/{2})(2H+c_{0})^{2}+\tilde{k}K+\tilde\lambda
+\tilde\gamma \kappa_{g}\right]_{C}=0\label{bound3}
\end{eqnarray}
are derived \cite{CapovillaPRE02,TuPRE03}. Here $\tilde{k}\equiv\bar{k}/k_c$ and $\tilde{\gamma}\equiv\gamma/k_c$ are the reduced bending modulus, and reduced line tension, respectively. $\kappa_n$, $\kappa_g$, and $\tau_g$ are the normal
curvature, geodesic curvature, and geodesic torsion of the boundary curve, respectively. The `dot' represents the derivative with respect to the arc length of the edge. Equation~(\ref{eq-openm}) expresses
the normal force balance of the membrane while equations (\ref{bound1})--(\ref{bound3}) represent the force and moment balances at each point in
curve $C$ \cite{CapovillaJPA02,TuJPA04}. Thus, in general, the above four
equations are independent of each other and available for an open membrane with several edges.

An axisymmetric surface can be generated by a planar curve $C_1C_2$ revolving around
an axis as shown in figure \ref{figgencuve}b. The above equations (\ref{eq-openm})--(\ref{bound3}) can be simplified as\cite{TuPRE03,TuJPA04}
\begin{eqnarray}
(h-c_{0})\left(\frac{h^{2}}{2}+\frac{c_{0}h}{2}-2K\right)-\tilde{\lambda}
h+\frac{\cos \psi }{\rho}(\rho\cos \psi h')'=0,\label{sequilib}
\\
\left[h-c_{0}+\tilde{k}{\sin \psi }/{\rho}\right]_C=0,\label{sbound1}\\
\left[-\sigma\cos \psi h'+\tilde{\gamma}{\sin \psi
}/{\rho}\right]_C=0,\label{sbound2}\\
\left[\frac{1}{2}(h-c_0)^2+\tilde{k}K+\tilde{\lambda}-\sigma\tilde{\gamma}
\frac{\cos \psi }{\rho}\right]_C=0,\label{sbound3}
\end{eqnarray}
where $C$ represents the edge point $C_1$ or $C_2$. $\sigma =1$ or $-1$ if the tangent vector $\mathbf{t}$ of the boundary curve is parallel or antiparallel to rotation direction respectively.

Similar to the above section, shape equation (\ref{sequilib}) is integrable, which can be reduced to a second order differential equation
\begin{equation}\cos\psi h'
+(h-c_{0}) \sin\psi\psi^{\prime}-\tilde{\lambda} \tan\psi+\frac{\eta_{0}}{\rho\cos\psi}-\frac{\tan\psi}
{2}(h-c_{0})^{2} =0\label{firstintg2}\end{equation} with an integral
constant $\eta_{0}$ \cite{TuJCP2010}. The configuration of an axisymmetric open lipid membrane should satisfy
shape equation (\ref{firstintg2}) and
boundary conditions (\ref{sbound1})--(\ref{sbound3}). In particular,
the points in the boundary curve should satisfy not only the
boundary conditions, but also shape equation (\ref{firstintg2})
because they also locate in the surface. That is,
equations~(\ref{sbound1})-(\ref{sbound3}) and (\ref{firstintg2}) should be
compatible with each other in the edge. Substituting
equations~(\ref{sbound1})-(\ref{sbound3}) into (\ref{firstintg2}), we
derive the compatibility condition \cite{TuJCP2010} to be
\begin{equation}\eta_{0}=0.\label{compat-cond}\end{equation}
Under this condition, the shape equation
is reduced to
\begin{equation}\cos\psi h'
+(h-c_{0}) \sin\psi\psi^{\prime}
  -\tilde{\lambda} \tan\psi-\frac{\tan\psi}%
{2}(h-c_{0})^{2} =0,\label{newshapeq}\end{equation}
while three boundary conditions are reduced to two equations, i.e. equations~(\ref{sbound1}) and (\ref{sbound3}).

\subsection{Finding solutions---mission impossible}

Now our task is to find analytic solutions that satisfy both the shape equation and the boundary conditions.
An obvious but trivial one is a planar circular disk with radius $R$. In this case, equations~(\ref{eq-openm})--(\ref{bound3}) degenerate into
\begin{equation}\tilde{\lambda} R+ \tilde{\gamma} =0.\end{equation}

\begin{figure}[pth!]
\centerline{\includegraphics[width=7cm]{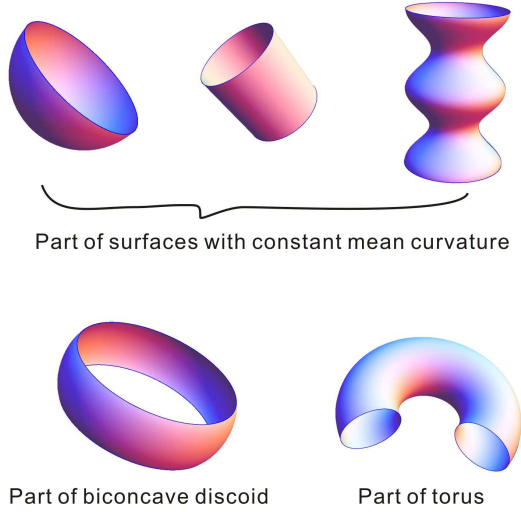}}
\caption{\label{fig-impopm}(color online)
Schematics of several impossible open membranes with free edges. Top: parts of sphere, cylinder and unduloid.
Bottom: parts of biconcave discoid and torus.}
\end{figure}

Can we find nontrivial analytic solutions? We have known some analytic solutions that satisfy the shape equation (\ref{eq-openm}), which include surfaces with constant mean curvature, biconcave discoid, torus and invert catenoid. Can we find a closed curve on these surface to satisfy the boundary conditions (\ref{bound1})--(\ref{bound3})? We will prove the following theorem of non-existence: For finite line tension, there does NOT exist an open membrane being a part of surfaces with constant (non-vanishing) mean curvature, biconcave discoid (valid for axisymmetric case), or Willmore surfaces (torus, invert catenoid). Several typical impossible open membranes with free edges are shown in figure~\ref{fig-impopm}.

The original version of this theorem was proposed in references~\citeonline{TuJGSP2011} and \citeonline{TuJCP2010}. Here we refine the original proof of this theorem and simultaneously correct some flaws.

Firstly, it is easy to prove there is no open membrane being a part of a spherical vesicle or cylindrical surface. The details are neglected here and they can be found in reference~\citeonline{TuJGSP2011}. We emphasize that the key obstacle happens in boundary condition (\ref{bound2}) which implies that the out-of-plane forces cannot balance in the edge.

Secondly, we will derive the second compatibility condition rather than (\ref{compat-cond}).
Let us consider the scaling transformation $\mathbf{r}\rightarrow (1+\epsilon)\mathbf{r}$, where the vector $\mathbf{r}$ represents the position of each point in the membrane and $\epsilon$ is a small parameter \cite{CapovillaJPA02,TuJGSP2011,TuJCP2010}. Under this transformation, we have $A\rightarrow (1+\epsilon)^2 A$, $L\rightarrow (1+\epsilon) L$, $H\rightarrow (1+\epsilon)^{-1} H$, and $K\rightarrow (1+\epsilon)^{-2} K$. Thus the free energy (\ref{eq-frenergyn2}) is transformed into $F_O (\epsilon)$.
The equilibrium configuration should satisfy $\partial F_O /\partial\epsilon =0$, from which we obtain the second compatibility condition
\begin{equation}2 c_0 \int  H \mathrm{d} A+(2\tilde\lambda+c_0^2) A +\tilde\gamma L=0.\label{constraitg}\end{equation}

Thirdly, we will prove there is no open membrane being a part of a curved surface with non-vanishing constant mean curvature. From the shape equation (\ref{eq-openm}), we derive $H=-c_0/2 \neq 0$ and $\tilde{\lambda}=0$ in this case, which contradict the compatibility condition (\ref{constraitg}) for $\gamma\neq 0$.

Fourthly, we will prove there is no axisymmetric open membrane being a
part of a biconcave discodal surface generated by a planar curve
expressed by $\sin\psi=c_0 \rho \ln(\rho/\rho_B)$. Substituting this equation into shape equation (\ref{firstintg2}),
we obtain $\tilde{\lambda}=0$ and $\eta_0 = -2 c_0\neq 0$ which contradicts to compatibility
condition (\ref{compat-cond}).

Finally, we consider the Willmore surface \cite{Willmore82} which satisfies the special form of equation~(\ref{eq-openm}) with vanishing $\tilde\lambda$ and $c_0$. Thus the compatibility condition (\ref{constraitg}) cannot be satisfied when $\tilde\lambda=0$ and $c_0=0$ because $\tilde\gamma L>0$. That is, there is no open membrane being a
part of Willmore surface which includes torus and invert catenoid.

Up to now, we have proven the theorem of non-existence, which implies that it is hopeless to find analytic solutions to the shape equation and boundary conditions of open lipid membranes. Thus the numerical simulations \cite{TuJCP2010,DuLWJCP06} are highly appreciated.

\subsection{Challenges}
Now we will discuss how we can further develop the above results on open lipid membranes.

\subsubsection{Minimal surface with boundary curve}
If carefully analyzing the above theorem and its proof, we will find that the minimal surface ($H=0$) is not touched. In fact, when $c_0=0$, $H=0$ with non-vanishing $\tilde{\lambda}$ can also satisfy the shape equation~(\ref{eq-openm}). Additionally, provided that $\tilde{\lambda}<0$, the minimal surface is consistent with compatibility
conditions (\ref{compat-cond}) and (\ref{constraitg}).

On the one hand, if $\tilde{k}$ is vanishing, the boundary condition (\ref{bound1}) holds naturally. Then boundary condition (\ref{bound2}) suggests $\kappa_n=0$. Further, boundary condition (\ref{bound3}) requires $\kappa_g =-\tilde{\lambda}/\tilde{\gamma}=\mathrm{constant}$.

On the other hand, if $\tilde{k}\neq 0$, the boundary condition (\ref{bound1}) gives $\kappa_n=0$. Then boundary condition (\ref{bound2}) suggests $\tau_g=\mathrm{constant}$. Since classical differential geometry tells us $K=-\tau_g^2$ when $\kappa_n=0$, boundary condition (\ref{bound3}) still requires $\kappa_g =\mathrm{constant}$.

In short, the big challenge is whether we can find a closed curve with vanishing normal curvature and constant geodesic curvature on some minimal surface except the planar circular disk.

\subsubsection{Neck condition of two-phase vesicles in the budding state}
\begin{figure}[pth!]
\centerline{\includegraphics[width=7cm]{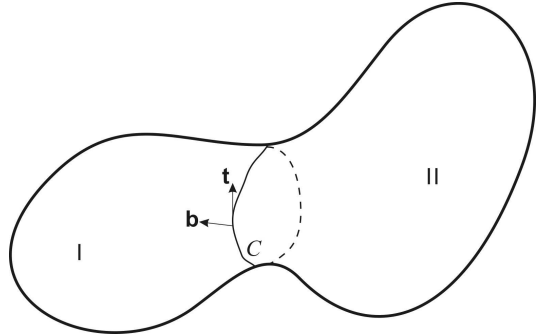}}
\caption{\label{fig-2phv}A vesicle with two phases (I and II) separated by curve $C$. $\mathbf{t}$ and $\mathbf{b}$ are located in the tangent plane of the surface. The former is the tangent vector of $C$ while the latter is perpendicular to $\mathbf{t}$ and points to the side of phase I.}
\end{figure}

The governing equations of open lipid membranes can be extended to a lipid vesicle with two phases separated by a boundary curve $C$ as shown in figure \ref{fig-2phv}.

The free energy of the two-phase vesicle can be expressed as
\begin{equation}F_{T}= \int_{\mathrm{I}} \left[ \frac{k_c^\mathrm{I}}{2} (2H +c_0^\mathrm{I})^2 + \bar{k}^\mathrm{I} K \right] \mathrm{d} A + \int_{\mathrm{II}} \left[ \frac{k_c^\mathrm{II}}{2} (2H +c_0^\mathrm{II})^2 + \bar{k}^\mathrm{II} K \right] \mathrm{d} A +\lambda^\mathrm{I} A^\mathrm{I}+ \lambda^\mathrm{II} A^\mathrm{II}+pV+\gamma L ,\label{eq-frenergy2pv}\end{equation}
where the superscripts indicate the mechanical parameters for each phase, for example, $c_0^\mathrm{I}$ and $c_0^\mathrm{II}$ are respectively the spontaneous curvatures for phase I and II.

Usually, we can derive the matching conditions that the curve $C$ should satisfy from the variation of the above free energy. But if noticing that the physical meanings of equations (\ref{bound1})--(\ref{bound3}) are the force or moment balances in the boundary, we can directly write down the matching conditions as follows\cite{TuJPA04}:
\begin{eqnarray}
&&\left[ k_c^\mathrm{I}(2H+c_{0}^\mathrm{I})-k_c^\mathrm{II}(2H+c_{0}^\mathrm{II})+(\bar{k}^\mathrm{I}-\bar{k}^\mathrm{II})\kappa_n\right]_{C} =0,\label{matchc1} \\
&&\left[\gamma
\kappa_n+(\bar{k}^\mathrm{I}-\bar{k}^\mathrm{II})\dot{\tau}_g-2(k_c^\mathrm{I}-k_c^\mathrm{II}){\partial H}/{\partial\mathbf{b}}\right]_{C} =0,\label{matchc2}\\
&&\left[\frac{k_c^\mathrm{I}}{2} (2H +c_0^\mathrm{I})^2-\frac{k_c^\mathrm{II}}{2} (2H +c_0^\mathrm{II})^2 + (\bar{k}^\mathrm{I}-\bar{k}^\mathrm{II}) K +(\lambda^\mathrm{I}-\lambda^\mathrm{II})+\gamma \kappa_{g}\right]_{C}=0.\label{matchc3}
\end{eqnarray}
We note that Das \textit{et al.} also obtained the equivalent form of above matching conditions in the axisymmetric case~\cite{Das2009}.

J\"{u}licher and Lipowsky investigated the budding of axisymmetric vesicles and found a limit shape
which is the state of two vesicles connected by a small neck. They also derived the neck condition\cite{Jul-lipowpre96,Seifert97ap}
\begin{equation}k_c^\mathrm{I}M^\mathrm{I}+k_c^\mathrm{II}M^\mathrm{II}=\frac{1}{2}\left[k_c^\mathrm{I}c_0^\mathrm{I}+k_c^\mathrm{II}c_0^\mathrm{II}+\gamma\right]\label{eq-JLneckcond}\end{equation}
without considering the Gaussian bending terms. Here $M^\mathrm{I}$ and $M^\mathrm{II}$ correspond to $-H^\mathrm{I}$ and $-H^\mathrm{II}$ for the points nearby the neck in domain I and II, respectively. They also conjectured that this neck condition holds for the asymmetric case and claimed the lack of a general proof to this conjecture\cite{KITPC2012}. It is not straightforward to drive the neck condition from the general matching conditions (\ref{matchc1})--(\ref{matchc3}), which is a challenge to be solved in the forthcoming years.

\section{Nonlocal theory of membrane elasticity}
There are two kinds of nonlocal theory of membrane elasticity. One is the area-difference elasticity, the other is the elasticity of membrane with nonlocal interactions between different points.

\subsection{Area-difference elasticity}
Since it is very difficult for lipid molecules to flip from one leaf to the other\cite{Sheetz-Singer}, when the membrane is bent from the planar configuration, the area of per lipid molecule in one leaf should be larger than the equilibrium value while the area of per lipid molecule in another leaf should be smaller than the equilibrium value. Considering the in-plane stretching or compression in each leaf, a nonlocal term $(k_r/2)(\int 2 H \mathrm{d}A)^2$ might be added to the free energy of membranes\cite{Evans1980,Svetina1985}. Here $k_r =k_a t^2/2A_0$ with $k_a$ and $t$ being the compression modulus and thickness of the monolayer, respectively, while $A_0$ is the prescribed area of the membrane. Considering this term, one might express the free energy of a vesicle as
\begin{equation}\label{eq-feADE1}
F_{AD}=\int\left[\frac{k_c}{2}(2H + c_0)^2 + \bar{k}K\right]\mathrm{d}A + \lambda A+ pV+\frac{k_r}{2}\left(\int 2H \mathrm{d}A\right)^2.
\end{equation}

Similarly, if the membrane is initially curved with (spontaneous) relative area difference $a_0$, the nonlocal term $(k_r/2)(\int 2H \mathrm{d}A+a_0)^2$ might be included in the free energy after the membrane is deformed\cite{miaoling1996}.
Thus the energy of a vesicle can be expressed as
\begin{equation}\label{eq-feADE2}
F_{ADE}=\int\left[\frac{k_c}{2}(2H + c_0)^2 + \bar{k}K\right]\mathrm{d}A + \lambda A+ pV+\frac{k_r}{2}\left(\int 2H \mathrm{d}A+a_0\right)^2.
\end{equation}
In fact, if we make a transformation $C_0=c_0+a_0k_r/k_c$ and $\Lambda=\lambda+a_0^2k_r/2A_0-k_ra_cc_0-k_r^2a_0^2/2k_c$, the above free energy is transformed into the form of equation (\ref{eq-feADE1}). Thus it is sufficient for us to consider the free energy (\ref{eq-feADE1}). The budding transitions of axisymmetric fluid-bilayer vesicles have been fully investigated on the basis of area difference elasticity~\cite{miaoling1996}. It is still necessary to discuss the general cases without presumption of axisymmetry.

\subsection{Membrane with nonlocal interactions}
Some lipid molecules contain charged head groups, thus molecules in different regions of membrane can interact with each other when two regions get close to each other. Intuitively, the free energy can be expressed as
\begin{equation}\label{eq-fenint}
F_{nint}=\int\left[\frac{k_c}{2}(2H + c_0)^2 + \bar{k}K\right]\mathrm{d}A + \lambda A+ pV+\varepsilon\int \mathrm{d}A\int \mathrm{d}A' U(|\mathbf{r}-\mathbf{r}'|),
\end{equation}
where $\mathbf{r}$ and $\mathbf{r}'$ represent the position vectors of different points in the membrane surface while $\mathrm{d}A$ and $\mathrm{d}A'$ are the area elements corresponding to the points $\mathbf{r}$ and $\mathbf{r}'$, respectively. $H$ and $K$ are the local mean curvature and Gaussian curvature at point $\mathbf{r}$, respectively. $\varepsilon$ and $U(.)$ represent the energy scale and the function form of nonlocal interactions, respectively.

Interestingly, we have proved that the helfrich bending energy (\ref{eq-helfrich}) with vanishing $c_0$ can be applicable to the bending of graphene \cite{Tu2008jctn,OYSUWangPRL97,TuOYPRB}. If we consider that $U(|\mathbf{r}-\mathbf{r}'|)$ is the Van der Waals-like interaction, the relative large camber arch\cite{LiuIijima2009,Yakobson09} in the edges of bilayer graphene might be understood on the basis of free energy (\ref{eq-fenint}) without osmotic pressure.

\subsection{Challenges}
Now we will discuss how we can further develop the above two kinds of nonlocal theory.

\subsubsection{Shape equation and its solutions to the shape equation of vesicles based on area-difference elasticity}
According to the variational method developed in our previous work\cite{TuPRE03,Tu2008jctn,TuJPA04}, the shape equation of vesicles which corresponds to the Euler-Lagrange equation of free energy (\ref{eq-feADE1}) can be derived as
\begin{equation}\tilde{p}-2\tilde{\lambda}
H+(2H+c_0)(2H^2-c_0H-2K)+\nabla^2(2H) -4\tilde{k}_rK\int H \mathrm{d}A=0\label{shape-adeclosed}\end{equation}
with reduced parameters $\tilde{p}=p/k_c$, $\tilde{\lambda}=\lambda/k_c$ and $\tilde{k}_r=k_r/k_c$. This is a fourth-order nonlinearly integro-differential equation, so it is hard for us to find some exact solutions to this equation.

Obviously, sphere is a solution to the above equation (\ref{shape-adeclosed}) which requires the radius $R$ of sphere satisfying
\begin{equation}\label{eq-sphade}
\tilde{p}R^2-(2\tilde{\lambda}+16\pi\tilde{k}_r)R-c_0(2-c_0R)=0.
\end{equation}
Comparing this equation with (\ref{sphericalbilayer}), we find that the nonlocal term has effect on the surface tension.

To check the other axisymmetric solutions, we adopt the representation shown in figure 1. In this representation, $2H=-h=-[{\sin \psi }/{\rho}+(\sin\psi)']$, $K={\sin \psi
}(\sin\psi)'/{\rho}$, $\nabla^2(2H)= -(\rho\cos \psi h')'\cos \psi/\rho$ and $\mathrm{d}A=2\pi|\sec\psi|\rho\mathrm{d}\rho$. Thus equation (\ref{shape-adeclosed}) is transformed into
\begin{equation}
\tilde{p}+\tilde{\lambda}
h+(c_{0}-h)\left(\frac{h^{2}}{2}+\frac{c_{0}h}{2}-2K\right)-\frac{\cos \psi }{\rho}(\rho\cos \psi h')'+4\pi \tilde{k}_r\left(\int h |\sec\psi|\rho\mathrm{d}\rho\right)K=0.\label{shapeade-symmetr}\end{equation}
It is necessary to note that the integral in the above equation is done on the minimal generation curve for the axisymmetric surface.

A torus can be generated by a planar curve expressed by (\ref{toruseqsym}). Substituting it into equation (\ref{shapeade-symmetr}), we still derive $R/r=\sqrt{2}$, while $2\tilde{\lambda}r=c_0(4-c_0r)-16\sqrt{2}\pi^2\tilde{k}_r r$ and $\tilde{p}r^2 =8\sqrt{2}\pi^2\tilde{k}_r r-2c_0$. That is, the torus with ratio of two generation radii being $\sqrt{2}$ is also the solution to the shape equation of vesicles within the framework of area difference elasticity.

Now we will check the biconcave surface generated by planar curve expressed by $\sin\psi =\alpha \rho \ln(\rho/\rho_B)$. We find that equation (\ref{shapeade-symmetr}) is satisfied when $\tilde{p}=0$, $\tilde{\lambda}=(\alpha^2 -c_0^2)/2$, $\alpha=c_0-\tilde{k}_r(4\pi|z_0|+\alpha A_0)$, where $z_0$ is the the coordinate of the pole shown in figure~\ref{figbiconcv} while $A_0$ represents the total area of the membrane. That is, the biconcave surface generated by planar curve expressed by $\sin\psi =\alpha \rho \ln(\rho/\rho_B)$ is also the solution to the shape equation of vesicles within the framework of area difference elasticity. Here the only difference is that $\alpha\neq c_0$ when we consider the area difference elasticity.

The three above examples imply that the shape equations of vesicles with and without consideration of the area difference elasticity seem to share the same form of solutions. Now we will verify this proposition is indeed true. Let us assume $c_0 = \bar{c}_0-2\tilde{k}_r\int H\mathrm{d}A$, then equation (\ref{shape-adeclosed}) is transformed into
\begin{equation}\tilde{p}-2\bar{\lambda}
H+(2H+\bar{c}_0)(2H^2-\bar{c}_0H-2K)+\nabla^2(2H)=0,\label{shp-adeclosedtran}\end{equation}
where $\bar{\lambda}=\tilde{\lambda}+(c_0^2-\bar{c}_0^2)/2$. The above equation has the same form as equation (\ref{shape-closed}), so the solutions to both equations have the same forms.
Therefore, here the challenge is the same as that proposed in section 2.3.

\subsubsection{Shape equation and its solutions to the shape equation of vesicles based on elasticity of membrane with nonlocal interactions}
According to the variational method developed in our previous work\cite{TuPRE03,Tu2008jctn,TuJPA04}, the shape equation of vesicles which corresponds to the Euler-Lagrange equation of free energy (\ref{eq-fenint}) can be derived as
\begin{equation}\tilde{p}-2\tilde{\lambda}
H+(2H+c_0)(2H^2-c_0H-2K)+\nabla^2(2H) + 2\tilde{\varepsilon}\int (U_\mathbf{n}-2HU) \mathrm{d}A^\prime=0,\label{shape-EMNI}\end{equation}
where $\tilde{\varepsilon}\equiv \varepsilon/k_c$, $U_\mathbf{n}=(\partial U/\partial R)\hat{R}\cdot \mathbf{n}$, $\mathbf{R}=\mathbf{r}^\prime-\mathbf{r}$, $R=|\mathbf{R}|$, $\hat{R}=\mathbf{R}/R$, $U=U(R)$. $\mathrm{d}A^\prime$ represents the area element at point $\mathbf{r}^\prime$. $H$, $K$ and $\mathbf{n}$ represent the mean curvature, the gaussian curvature and normal vector of membrane at point $\mathbf{r}$, respectively.

Since the nonlocal term $\int (U_\mathbf{n}-2HU) \mathrm{d}A^\prime$ depends on the vector $\mathbf{r}$ for given function form of $U$, this term is equivalent to a nonuniform pressure applied on the membrane. Sphere is an obvious solution to equation (\ref{shape-EMNI}) because the nonlocal term $\int (U_\mathbf{n}-2HU) \mathrm{d}A^\prime$ gives a constant quantity which corresponds to a uniform pressure. Thus equation (\ref{shape-EMNI}) still reduces to the same form of equation (\ref{sphericalbilayer}) which determines the radius of the sphere.
It is quite complicated to find the solutions corresponding to the shapes rather than spheres because the nonlocal term depends not only on the position of point in the membrane surface, but also on the function form of $U$. In particular, presuming $U$ to be the Van der Waals-like form, can we find some solutions rather than spherical shape?

\section{Relationship between symmetry and the magnitude of free energy}
Lipid vesicles in homogenous phase observed in experiments usually have higher degrees of symmetry such as spherical or axial symmetry \cite{HotaniJMB84}. In theoretical researches, most of vesicles are assumed to be axisymmetrical. Scientists seem to believe that the vesicles correspond to lower free energy if they have the higher degrees of symmetry under the same external conditions. To what extent this insight is true?

\subsection{Symmetry and symmetry broken viewed from the free energy}
\begin{figure}[pth!]
\centerline{\includegraphics[width=7cm]{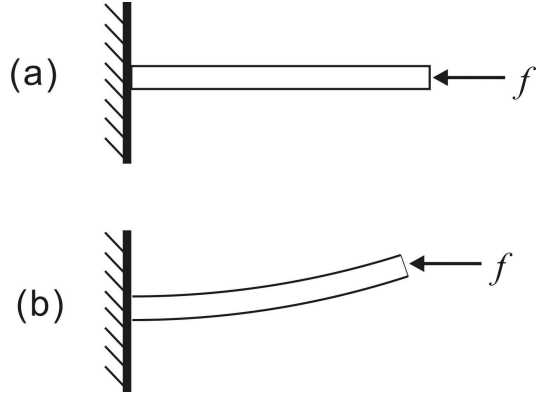}}
\caption{\label{fig-symbrk}Schematic of symmetry broken: (a) Straight conformation of a rod; (b) Bent conformation of a rod.}
\end{figure}

There exists some relationship between symmetry and the free energy of a structure. Let us consider a classic example shown in figure~\ref{fig-symbrk}. An external force $f$ is applied along the axis of a long elastic rod. Assume the centerline of the rod is inextensible, so the only mode of deformation is the deflection of the rod. Assume that the centerline of bent rod can be regarded as an arc of a circle with radius $R$. Note that this assumption is not accurate, while it can help us qualitatively and semi-quantitatively understand the main insights. The length of the rod is $L$, so the corresponding angle made by the arc can be expressed as $\theta=L/R$.

\begin{figure}[pth!]
\centerline{\includegraphics[width=7cm]{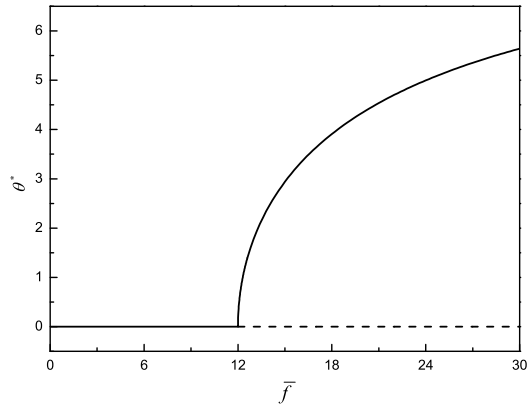}}
\caption{\label{fig-thetaforce}Numerical solutions to equation (\ref{eq-BuckleE2}) for various values of $\bar{f}$. There is only one solution ($\theta^\ast=0$, the solid line) when $\bar{f}<12$ while two solutions ($\theta^\ast=0$, the dash line; and $\theta^\ast >0$, the solid line) when $\bar{f}>12$.}
\end{figure}

The free energy of the system can be expressed as\cite{RobPhl}
\begin{equation}
\label{eq-BuckleE} F_r = {k_b \over L}
{\theta^2\over 2} - {f L} \left( 1 -
{2\over\theta}\sin{\theta\over 2} \right ),
\end{equation}
where $k_b$ is the bending rigidity of the rod. From $\partial F_r /\partial \theta =0$, we derive
\begin{equation}
\label{eq-BuckleE2} \theta^3 +\bar{f} \left(\theta \cos{\theta\over 2} -
2\sin{\theta\over 2} \right )=0,
\end{equation}
where $\bar{f}\equiv fL^2/k_b$ is the reduced force. We numerically solve the above equation, and then find the bifurcate behavior of the solutions which is shown in figure~\ref{fig-thetaforce}. There is only one solution ($\theta^\ast=0$, the solid line) when $\bar{f}<12$ while two solutions ($\theta^\ast=0$, the dash line; and $\theta^\ast >0$, the solid line) when $\bar{f}>12$.

\begin{figure}[pth!]
\centerline{\includegraphics[width=7cm]{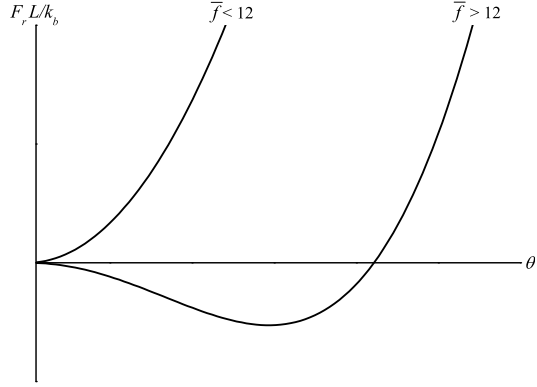}}
\caption{\label{fig-febrk}Typical diagrams of the relationship between the reduced free energy ($F_r L/k_b$) and the angle $\theta$.}
\end{figure}

The solution $\theta^\ast=0$ corresponds to the straight configuration while $\theta^\ast >0$ represents the bent configuration. Which one is in favor of lower free energy? In figure~\ref{fig-febrk}, we draw the typical diagrams of the relation between the reduced free energy ($F_r L/k_b$) and the angle $\theta$ for $\bar{f}<12$ and $\bar{f}>12$, respectively. We readily see that there is only one stationary point at $\theta^\ast =0$ which makes $\mathrm{d}F_r/\mathrm{d}\theta=0$ when $\bar{f}<12$, and this point also makes the free energy to take minimum value. On the other hand, when $\bar{f}>12$ there are two stationary points which make $\mathrm{d}F_r/\mathrm{d}\theta=0$. One point is located at $\theta^\ast =0$ which corresponds to a local maximum of the free energy; the other point is located at $\theta^\ast >0$ which corresponds to a local minimum of the free energy. In particular, the latter point is in favor of the free energy when $\bar{f}>12$.
Based on the above analysis, we find that the bent configuration (with lower symmetry) is in favor of lower free energy for larger compression force while the straight one (with higher symmetry) is in favor of lower free energy for smaller compression force. Of course, in the extreme case, the straight configuration (with higher symmetry) is always in favor of lower free energy for stretching force. Thus there is certain relationship between symmetry and free energy under specific conditions.

\subsection{Challenge---a conjecture}
In fact, the experimental results \cite{HotaniJMB84} also reveal that there is certain relationship between symmetry of the shape and free energy of a vesicle. Under lower osmotic pressure, the biconcave discoidal vesicle is of axial symmetry. Under the higher osmotic pressure, the vesicle is transformed into triangle like ($C_3$ symmetry) or even into other nonsymmetric shapes. Combining these experimental observations and the analysis on elastic rod, we conjecture: for the given area, the spherically topological vesicle with higher symmetry corresponds to lower Helfrich free energy (\ref{eq-exthelfrich}) if the osmotic pressure is small enough.

We can verify this conjecture for nearly spherical vesicle with zero excess area and small excess volume~\cite{OYbook1999}. A spherical vesicle with radius $R$ can be expressed as vector form $R\hat{R}$ where $\hat{R}$ represents the unit radial vector. The nearly spherical vesicle can be expressed as $\mathbf{r}=R[1+\sum_{lm} a_{lm}Y_{lm}(\theta,\phi)]\hat{R}$ with $|a_{lm}|\ll 1$, where $Y_{lm}(\theta,\phi)$ is the spherical harmonics satisfying $Y_{00}=1/\sqrt{4\pi}$ and $\nabla^2 Y_{lm}=-l(l+1)Y_{lm}$. Then the excess area can be expressed as
\begin{equation}A_{ex}=A_\mathrm{vesicle}-A_\mathrm{sphere}=4\sqrt{\pi}a_{00}R^2+{1\over 2}\sum_{lm} [l(l+1)+2]|a_{lm}|^2 R^2=0\label{eq-exsA1}\end{equation}
up to the second order term of $a_{lm}$.
Similarly, the excess volume can be expressed as
\begin{equation}V_{ex}=V_\mathrm{vesicle}-V_\mathrm{sphere}=2\sqrt{\pi}a_{00}R^3+R^3\sum_{lm}|a_{lm}|^2.\label{eq-exsV1}\end{equation}
The bending energy can be expressed as
\begin{equation}F_c= 2\pi k_c(2-c_0R)^2- 2k_c \sqrt{\pi}a_{00}c_0R(2-c_0R)+{k_c \over 2}\sum_{lm}|a_{lm}|^2[l^2(l+1)^2-l(l+1)(2+2c_0R-c_0^2R^2/2)+c_0^2R^2]\end{equation}
Then the Helfrich free energy can be expressed as $F=F_c+ \lambda (4\pi R^2+ A_{ex})+ p (4\pi R^3/3+V_{ex})$.
When $|a_{lm}|\ll 1$, minimizing $F$ with respect to $a_{00}$, we derive $ \lambda =k_c c_0(2-c_0R)/2R-pR/2$. Substituting it into the expression of $F$, we obtain
\begin{equation}F= F_\mathrm{sphere}+{k_c \over 2}\sum_{lm}|a_{lm}|^2[l(l+1)-2][l(l+1)-c_0R-pR^3/2k_c],\end{equation}
where $F_\mathrm{sphere}=4\pi k_c(2-c_0R)-2\pi pR^3 /3$ is the free energy of the sphere.

The $Y_{00}$ mode cannot be excited separately because of the constraint (\ref{eq-exsA1}). The $Y_{1m}$ mode is trivial, which represents the small translation of the sphere. If $p<2(6-c_0R)/R^3$, the all excited $Y_{lm}$ modes make $F>F_\mathrm{sphere}$, i.e., increase the free energy. Thus the spherical shape (the higher symmetry) corresponds to lower free energy among all nearly spherical vesicles. However, it is a big challenge to prove this conjecture globally for larger excess volume.

\section{Conclusion}
In the above discussions, we present some key results in the theoretical investigations on configurations of lipid membranes. We also propose several challenges in this field, which are specifically highlighted again as follows.

\textit{Challenge 1}. Can we further find analytic solutions rather than sphere, torus and biconcave discoid to the shape equation (\ref{shape-closed}) or (\ref{firstintg}) which represent the closed vesicles without self-contact? An alternative scheme is to find solutions to equations (\ref{eq-varxi}) and (\ref{eq-simpshe}) rather than the original shape equation. If all these efforts are in vain, can we verify among all closed non-intersect surfaces, there are only sphere, torus and biconcave discoid that can satisfy the shape equation and can be expressed as the elementary functions?

\textit{Challenge 2}. Can we find a closed curve with vanishing normal curvature and constant geodesic curvature on some minimal surface except the planar circular disk? Or else, can we prove that the planar circular disk is the unique minimal surface with boundary curve which has vanishing normal curvature and constant geodesic curvature?

\textit{Challenge 3}. Can we drive the neck condition (\ref{eq-JLneckcond}) from the general matching conditions (\ref{matchc1})--(\ref{matchc3})?

\textit{Challenge 4}. Can we find the solutions rather than sperical shapes to the shape equation (\ref{shape-EMNI}) on the basis of elasticity of membrane with nonlocal Van der Waals-like interactions?

\textit{Challenge 5}. Can we prove the conjecture that among all spherically topological vesicles the configuration with higher symmetry corresponds to lower Helfrich free energy (\ref{eq-exthelfrich}) if the osmotic pressure is small enough?

Researchers have made fruitful achievements in the field of membrane biophysics since 1970s. These achievements have also gained much recognition in the scientific community. In 2012, Helfrich was awarded the Charles Stark Draper Prize for the engineering development of the liquid crystal display utilized in billions of consumer and professional devices, and the Raymond and Beverly Sackler International Prize in Biophysics for his contributions to the biophysics of lipid bilayers and biological membranes. J\"{u}licher was awarded the 2007 Raymond and Beverly Sackler International Prize in Biophysics for his seminal contributions to the field of the physics of non-equilibrium bio-cellular systems such as molecular motors, active membranes, filaments and the cytoskeleton. With the increasing maturity of theoretical investigations on biological membranes, the remained problems are more difficult than before. Among them, I believe that the above five challenges are very significant for theoretical investigations on configurations of biological membranes and they are highly expected to be overcome through the collaborations between mathematicians and physicists in the forthcoming years.

\section*{Acknowledgement}
The author is grateful to professor Zhong-can Ou-Yang for his kind suggestions. He also thanks Pan Yang and Yang Wang for their carefully proofreading the manuscript.

\end{CJK*}  
\end{document}